\documentclass[fleqn,10pt]{wlscirep}
\usepackage{amssymb}
\usepackage{amsmath}
\usepackage{graphicx}
\usepackage{epsfig}
\usepackage{graphics}
\usepackage{subfigure}

\title{Magnetic-flux-driven topological quantum phase transition and
manipulation of perfect edge states in graphene tube}

\author[1]{S. Lin }
\author[2]{ G. Zhang }
\author[1]{ C. Li }
\author[1,*]{ Z. Song}
\affil[1]{School of Physics, Nankai University, Tianjin 300071, China}
\affil[2]{College of Physics and Materials Science, Tianjin Normal University, Tianjin
300387, China}
\affil[*]{songtc@nankai.edu.cn}
\begin{abstract}
We study the tight-binding model for a graphene tube with perimeter $N$
threaded by a magnetic field. We show exactly that this model has different
nontrivial topological phases as the flux changes. The winding number, as an
indicator of topological quantum phase transition (QPT) fixes at $N/3$\ if $
N/3$ equals to its integer part $\left[ N/3\right] $, otherwise it jumps
between $\left[ N/3\right] $ and $\left[ N/3\right] +1$ periodically as the
flux varies a flux quantum. For an open tube with zigzag boundary condition,
exact edge states are obtained. There exist two perfect midgap edge states,
in which the particle is completely located at the boundary, even for a tube
with finite length. The threading flux can be employed to control the
quantum states: transferring the perfect edge state from one end to the
other, or generating maximal entanglement between them.
\end{abstract}

\begin{document}

\maketitle

\flushbottom

\section*{Introduction}

As a two-dimensional carbon sheet of single-atom thickness, graphene has been used
as one kind of promising material for many aspects of display screens, electric circuits, and solar cells, as
well as various medical, chemical and industrial processes \cite%
{Pauling,Saito,Novoselov1,Charlier,Abergel,Cooper}. Due to its special
structure, graphene possesses a lot of unique properties in chemistry,
physics and mechanics \cite%
{Diankov,Semenoff,Gusynin,Novoselov2,Zhang1,Katsnelson,Zhang2,
Avouris,Akhmerov,Chen,Steinberg,Neto,Lamas,Jobst,Alexander-Webber,Baringhaus,Carlsson,Lee}
. Recently, 3D graphene materials have received much attention, since they
not only possess the intrinsic properties of 2D graphene sheets, but also
provide the advanced functions with improved performance in various
applications \cite{Geim,Matyba,Liu1,Liu2,Yankowitz,Dauber}.

Between the two regimes, a geometrical tube system, as a quasi-3D graphene,
is expected to exhibit novel properties, especially under the control of
external degrees of freedom. In this paper, we theoretically investigate a
graphene tube threaded by a magnetic field in the tight-binding framework.
We show exactly that a honeycomb lattice tube can be reduced to two
equivalent Hamiltonians with a tunable distortion by flux, one of which is a
combination of several independent dimerized models. And the other one is a
1D system with long-range hoppings. The flux drives the QPTs with
multi-critical points. Applying the geometrical representation on the two
equivalent Hamiltonians, we find that, although two loops show different
geometries, they always have the same topology, i.e., as the flux varies the
winding number around a fixed point can be the indicator of topological QPT.
We also investigate the zero modes for an open tube with zigzag boundary
condition by exact Bethe ansatz solutions. We find that there exist two
perfect edge states at the midgap, in which the particle is completely
located at the boundary, when a proper flux is applied. Remarkably, such
zero modes still appear in a tube with small length, which allows us to
design a nanoscale quantum device. Using the threading flux as an external
control parameter, the perfect edge state can be transferred from one end to
the other, or the maximal entanglement between them can be generated by an
adiabatic process. 

\begin{figure}[tbp]
\centering
\includegraphics[ bb=71 182 482 618, width=0.47\textwidth, clip]{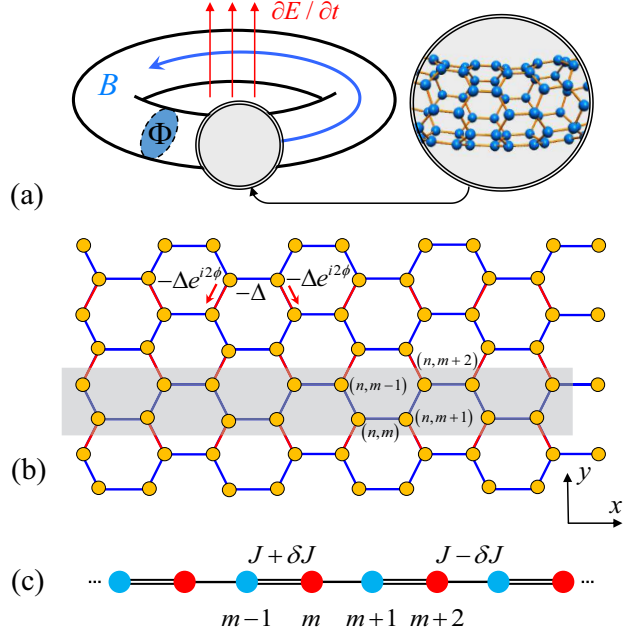}
\par
\caption{(color online). (a) Sketch of a graphene tube threaded by a
magnetic field considered in this work. The flux induced by an external
varying electric field or current, derives the topological QPTs and quantum
state transfer. (b) Honeycomb lattice of a tube with periodic boundary
condition. The tube shows translational symmetry along $x$ and $y$ axes. The
complex hopping constants induced by the flux are denoted in red. The gray
shaded region represents a lattice ring as the super unit cell of the tube,
which repeats itself along $y$ axis. (c) Sketch of a SSH lattice ring
reduced from the tube. The hopping integral $J$\ and dimerization strength $%
\protect\delta $\ are the functions of flux.}
\label{fig1}
\end{figure}

\section*{Results}

\vspace{2ex}\textbf{Model and solutions.} The tight-binding model for a
honeycomb tube in the presence of a threading magnetic field can be
described by the Hamiltonian%
\begin{equation}
H=-\Delta \sum_{n=1}^{N}(\sum_{m=1}^{M}c_{n,m}^{\dagger }c_{n,m+1}+e^{i2\phi
}\sum_{m=1}^{M/4}c_{n,4m}^{\dagger }c_{n+1,4m-1}+e^{i2\phi
}\sum_{m=1}^{M/4}c_{n,4m-3}^{\dagger }c_{n+1,4m-2}+\mathrm{H.c.}),  \label{H}
\end{equation}%
where $c_{n,m}$ ($c_{n,m}^{\dagger }$) annihilates (creates) an electron at
site $\left( n,m\right) $ on an $N\times M$ lattice\ with integer $M/4$ and $%
N\geqslant 3$, and obeys the periodic boundary conditions, $%
c_{n,M+1}=c_{n,1} $, $c_{N+1,4m-1}=c_{1,4m-1}$, and $c_{N+1,4m-2}=c_{1,4m-2}$%
, with $n\in \left[ 1,N\right] $, $m\in \left[ 1,M/4\right] $. Parameter $%
\Delta $ is the hopping integral with zero magnetic field \textbf{(}we take $%
\Delta =1$ hereafter for simplification\textbf{)} and $\phi =\pi \Phi
/(N\Phi _{0})$, where\ $\Phi $\ is the flux threading the tube, $\Phi _{0}$
is flux quantum. In Fig. \ref{fig1}, the geometry of the model is
illustrated schematically. We are interested in the effect of $\phi $\ on
the property of the system with small $N$ and large $M$. To this end, a
proper form of solution is crucial. We employ the Fourier transformation $%
a_{K,l}^{\dagger }=\frac{\eta _{l}}{\sqrt{N}}\sum_{j=1}^{N}e^{iKj}c_{j,l}^{%
\dagger }$, with%
\begin{equation}
\eta _{l}=\left\{
\begin{array}{cc}
1, & \left( l=4m,4m-3\right) \\
e^{-i\left( K/2+\phi \right) }, & \left( l=4m-1,4m-2\right)%
\end{array}%
\right. ,
\end{equation}%
to rewrite the Hamiltonian, where $m\in \left[ 1,M/4\right] $, and $K=2\pi
n/N$, $n\in \left[ 1,N\right] $. The Hamiltonian can be expressed as $%
H=\sum_{K}H_{K}$, where%
\begin{equation}
H_{K}=-J\sum_{m=1}^{M/2}[\left( 1+\delta \right) a_{K,2m-1}^{\dag
}a_{K,2m}+\left( 1-\delta \right) \sum_{m=1}^{M/2}a_{K,2m}^{\dag }a_{K,2m+1}+%
\mathrm{H.c.}],  \label{H_K}
\end{equation}%
with periodic boundary $a_{K,M+1}=a_{K,1}$. Together with $\left[
H_{K},H_{K^{\prime }}\right] =0$, we find that $H$ is a combination of $N$
independent Peierls rings with the ($K,\phi $)-dependent hopping integral $%
J\left( K,\phi \right) $ and dimerization strength $\delta \left( K,\phi
\right) $, where $\delta =1-1/J$ and $J=\cos \left( K/2+\phi \right) +1/2$.
\begin{figure}[tbp]
\centering
\includegraphics[ bb=50 137 518 584, width=0.5\textwidth, clip]{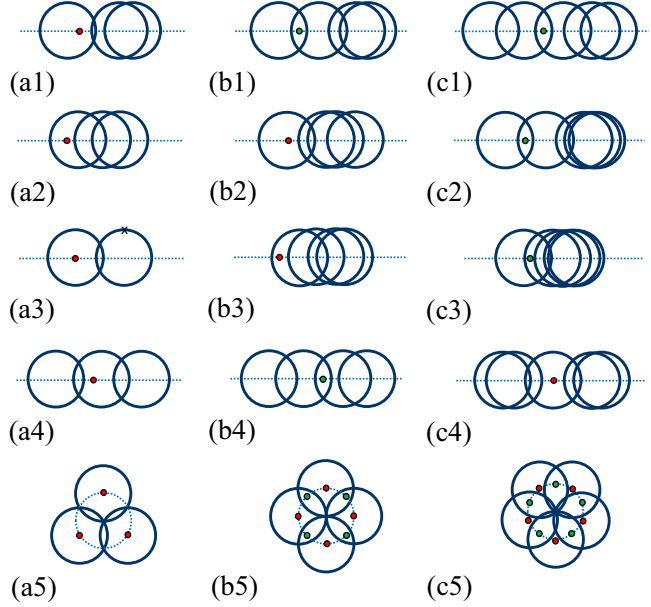}
\par
\caption{(color online). Several typical graphs in the parameter space for
the graphene tubes with $N=3,4,$ and $5$. The graphs are $N$ circles with
unit radius. The flux determines the positions of the circle centers. The
red and green circles indicate the positions of the origin. The
corresponding winding numbers are $[N/3]\ $or$\ [N/3]+1$, when the origin
circles are red or green. (a1-4), (b1-4), and (c1-4) are obtained by the
first equivalent Hamiltonian in Eq. (\protect\ref{H}), while (a5), (b5), and
(c5) by the second equivalent Hamiltonian in Eq. (\protect\ref{H_eq}). The
cross in (a2) indicates two superposed circles.}
\label{fig2}
\end{figure}
The one-dimensional dimerized Peierls system at half-filling, proposed by
Su, Schrieffer, and Heeger (SSH) to model polyacetylene \cite{Su,Schrieffer}%
, is the prototype of a topologically nontrivial band insulator with a
symmetry protected topological phase \cite{Ryu,Wen}. In recent years, it has
been attracted much attention and extensive studies have been demonstrated
\cite{Xiao,Hasan,Delplace,ChenS1,ChenS2}. For simplicity, $H_{K}$\textbf{\ }%
can be expressed as
\begin{equation}
H_{K}=-\sum_{m=1}^{M/2}(\kappa a_{K,2m-1}^{\dag }a_{K,2m}+a_{K,2m}^{\dag
}a_{K,2m+1})+\mathrm{H.c.},  \label{H_K2}
\end{equation}%
by defining\textbf{\ }$\kappa =J\left( 1+\delta \right) $. The topologically
nontrivial phase for the Hamiltonian $H_{K}$ with even $M$\ is realized for $%
\left\vert \kappa \right\vert <1$. It can only be transformed into a
topological trivial phase by either breaking the symmetries which protect it
or by closing the excitation gap. We note that $\kappa $\ is the function of
$\phi $, which indicates the flux can drive a topological QPT. The present
SSH ring described by $H_{K}$\ could lead to two transition points at $%
\kappa =\pm 1$, respectively, which will be seen from another point of view
in the next section.

\vspace{2ex}\textbf{Topological characterizations.} Now we investigate the
QPT from another perspective. For each SSH ring, we can represent $H_{K}$\
in a simple form%
\begin{equation}
H_{K}=2\sum_{k}\vec{r}\left( K,k\right) \cdot \vec{s}_{K,k},
\end{equation}%
where $\vec{s}_{K,k}$ is defined as pseudo spin operators $%
s_{K,k}^{+}=(s_{K,k}^{-})^{\dag }=a_{K,k}^{\dag }b_{K,k}$ and $s_{K,k}^{z}=%
\frac{1}{2}(a_{K,k}^{\dag }a_{K,k}-b_{K,k}^{\dag }b_{K,k})$, which obey the
Lie algebra relations $[s_{K,k}^{+},s_{K,k^{^{\prime
}}}^{-}]=2s_{K,k}^{z}\delta _{k,k^{^{\prime }}}$ and $%
[s_{K,k}^{z},s_{K,k^{^{\prime }}}^{\pm }]=\pm s_{K,k}^{\pm }\delta
_{k,k^{^{\prime }}}$. Here the fermion operators in $k$ space are $a_{K,k}=%
\sqrt{2/M}\sum_{m=1}^{M/2}e^{-ikm}a_{K,2m-1}$ and $b_{K,k}=\sqrt{2/M}%
\sum_{m=1}^{M/2}e^{-ikm}a_{K,2m}$ with $k=4\pi n/M$, $n\in $ $\left[ 1,M/2%
\right] $, and the components of the field $\vec{r}\left( K,k\right) $ in
the rectangular coordinates are%
\begin{equation}
\left\{
\begin{array}{c}
x\left( k\right) =-x_{0}-\cos k, \\
y\left( k\right) =-\sin k,%
\end{array}%
\right. ,  \label{curve1}
\end{equation}%
with $x_{0}=\kappa $. According to the analysis in Ref. \cite{ZG}, the
quantum phase has a connection to the geometry of the curves with parameter
equation $\left\{ x\left( k\right) ,y\left( k\right) \right\} $.
Specifically, the phase diagram can be characterized by the winding number
of the loop around the origin. We note that the loop of Eq. (\ref{curve1})
represents a circle with unit radius and the center of the circle is $\left(
K,\phi \right) $-dependent, i.e., $x_{0}=2\cos \left( K/2+\phi \right) $.
For any given $K$, as the flux $\phi $\ changes, $x_{0}$ varies within the
region $\left[ -2,2\right] $. When $x_{0}=\pm 1$, the circle crosses the
origin, leading to the switch of the winding number of the loops around the
origin. Since a given honeycomb tube consists of a set of SSH rings, the
topology of the corresponding circles reflects the feature of the system. A
straightforward analysis shows that the winding number $\nu $ depends on $N$
and flux $\phi $\ in a simple form:%
\begin{equation}
\nu =\left\{
\begin{array}{cc}
N/3, & N/3=\left[ N/3\right] \\
\left[ N/3\right] +\Lambda , & \left( N-1\right) /3=\left[ N/3\right] \\
\left[ N/3\right] +1-\Lambda , & \left( N-2\right) /3=\left[ N/3\right]%
\end{array}%
\right. ,  \label{WN}
\end{equation}%
where $\left[ x\right] $\ stands for the integer part of $x$ and $\Lambda
=[\left\vert 2\sin \left( N\phi \right) /\sqrt{3}\right\vert ]$. It
indicates that the topology of the tube state is unchanged if $N$ is a
multiple of $3$, or jumps between two cases, as $\phi $\ varies. In Fig. \ref%
{fig2}, we demonstrate the relationship between $\nu $\ and the geometry of
the loops for several typical $N$.
\begin{figure*}[tbp]
\centering
\includegraphics[ bb=77 236 501 540, width=0.32\textwidth, clip]{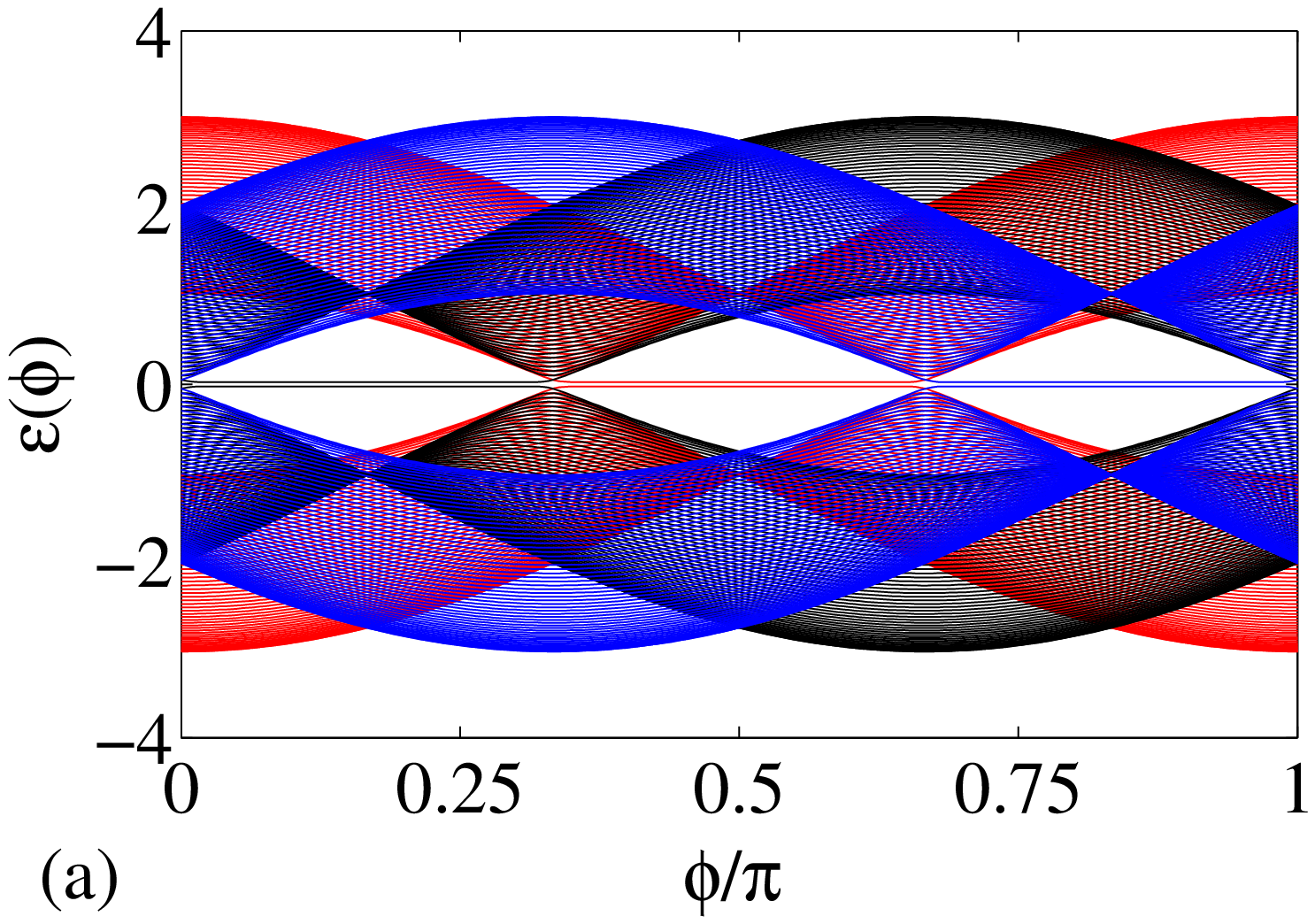} %
\includegraphics[ bb=77 236 501 540, width=0.32\textwidth, clip]{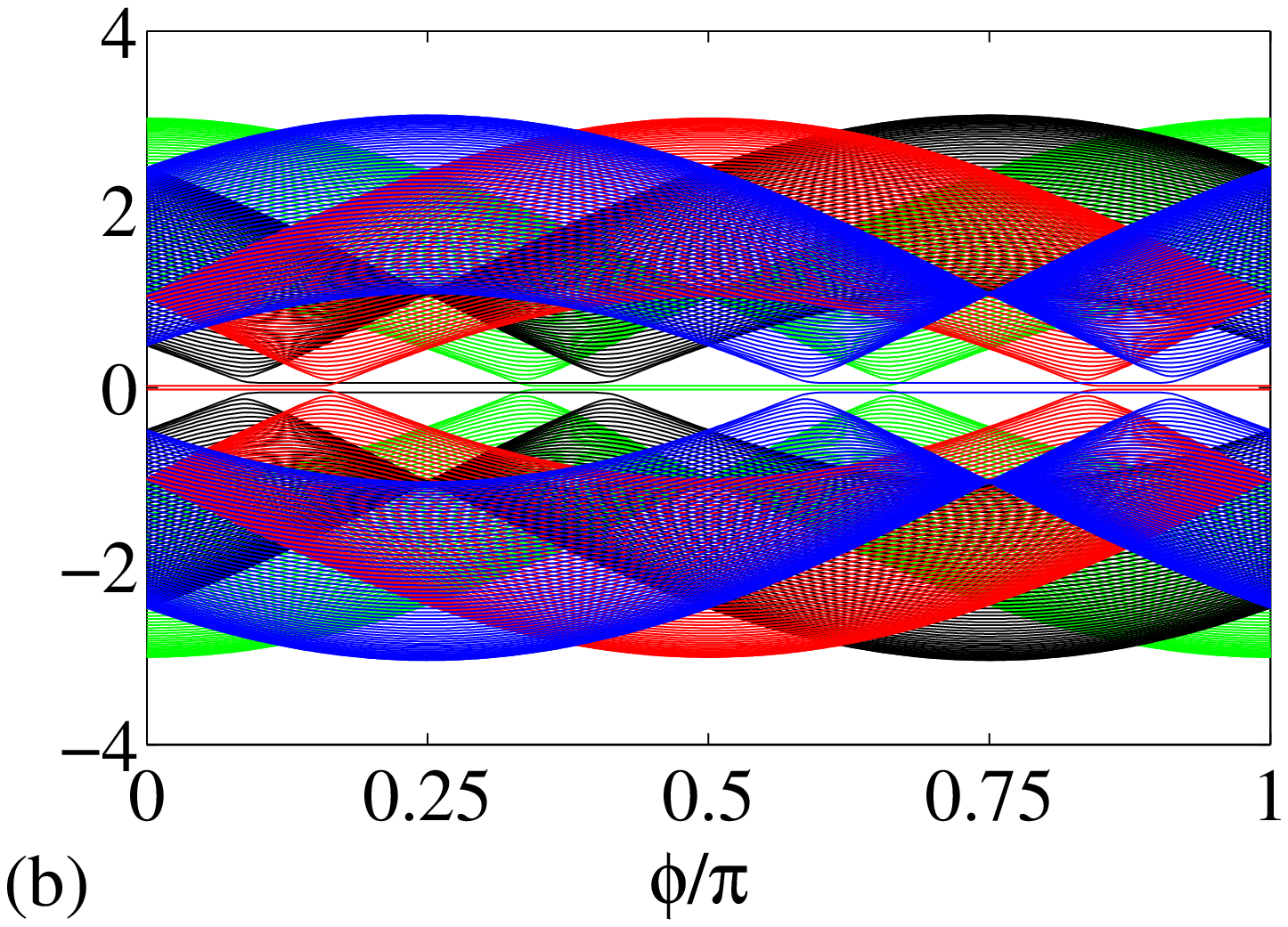} %
\includegraphics[ bb=77 236 501 540, width=0.32\textwidth, clip]{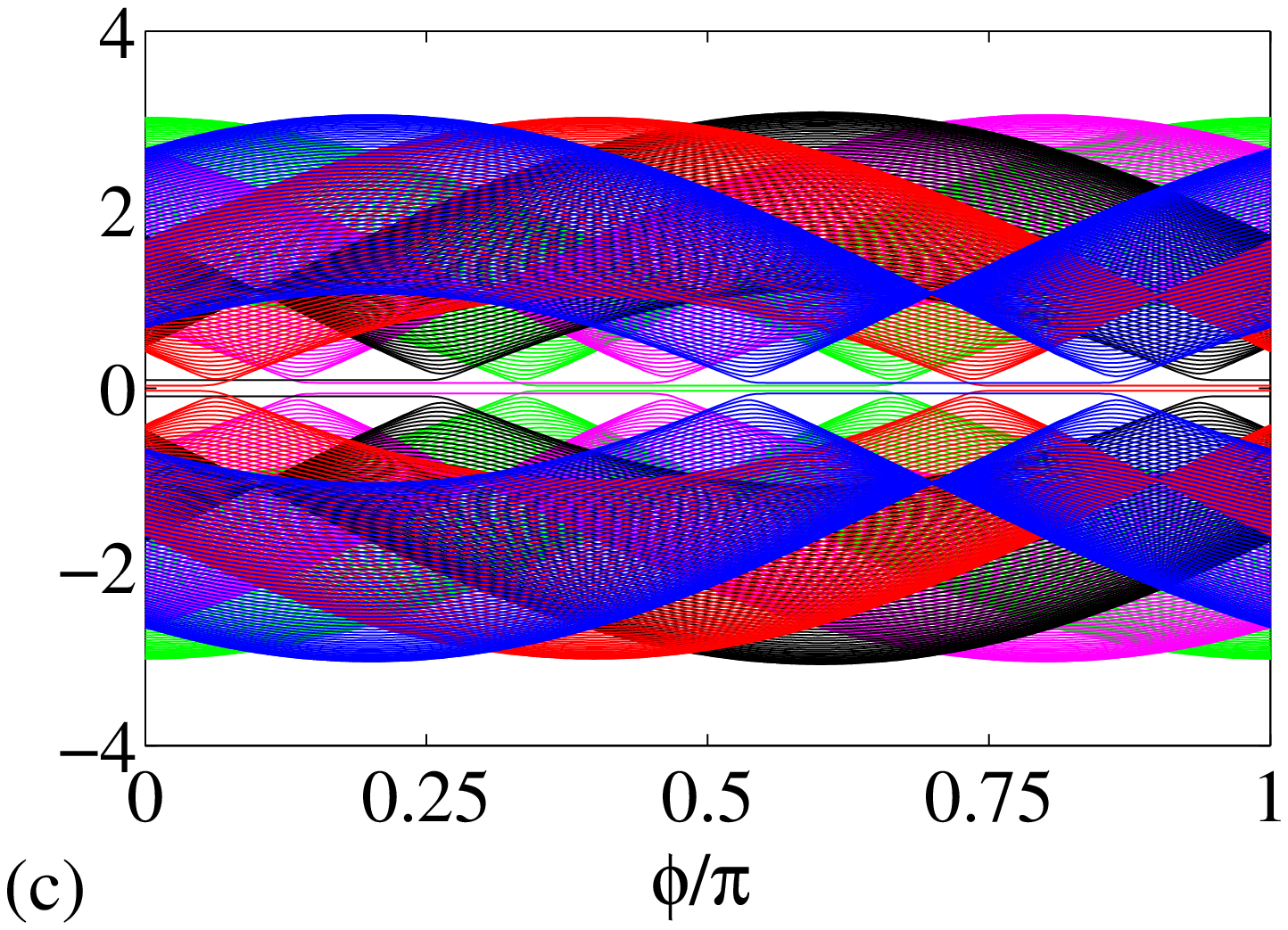}
\par
\caption{(color online). Energy spectrum of the finite-sized graphene tube
(a) $N=3,M=124$, (b) $N=4,M=124$, and (c) $N=5,M=124$, with flux $\protect%
\phi $. In order to present the zero modes clearly, we separate the upper
and lower bands by shifting them up and down with a small amount of energy,
respectively. It shows that the zero modes appear and disappear periodically
as the flux changes. Here the energy spectrum $\protect\varepsilon (\protect%
\phi )$ is expressed in units of the hopping constant $\Delta $.}
\label{fig3}
\end{figure*}
We have seen that the quantum phases can be characterized by the topology of
the loops in the parameter space. It seems that the obtained result is based
on the Fourier transformation. However, the theory of topological insulator
claims that the topological character is independent of the representation
and can be observed in experiment. To demonstrate this point, we consider
another representation, which maps the original Hamiltonian to a single ring
but with long-range hoppings and is shown in detail in Methods section. The
geometrical representation is clear, still consisting of $N$ cycles with
unit radius. The centers of these circles locate at another unit-radius
circle with equal distance, called circle-center circle. As flux varies, the
origin of the parameter space moves along the circle-center circle,
resulting in various values of winding numbers. A straightforward analysis
shows that although the pattern is different, it presents the same topology
with the first representation and Fig. \ref{fig2} illustrates the patterns
for small $N$.

\vspace{2ex}\textbf{Control of perfect edge states.} The above results
indicate that the quantum phase of the model $H$ exhibits topological
characterization. Another way to unveil the hidden topology behind the model
is exploring the zero modes of the system with open boundary condition.
Consider the graphene system with zigzag boundary and its Hamiltonian could
be rewritten as
\begin{equation}
H_{\mathrm{open}}=H+\sum_{n=1}^{N}(c_{n,1}^{\dagger }c_{n,M}+\mathrm{H.c.}).
\end{equation}%
Performing the Fourier transformation, $H_{\mathrm{open}}$\ can be
decomposed into $N$ independent SSH chains. The number of zero modes is
determined by the sign of $\left\vert \kappa \right\vert -1$, which leads to
the same conclusion as that from the above two geometrical representations.
In Fig. \ref{fig3} we plots the band structures for the systems demonstrated
in Fig. \ref{fig2}. It clearly demonstrates the processes of the emergence
and disappearance of zero modes. According to the Bethe ansaz results (see
Methods section.), the exact wave functions of edge states for a finite $N$
but infinite $M$ tube can be expressed as%
\begin{equation}
\left(
\begin{array}{c}
\langle j,l\left\vert \psi _{\mathrm{L}}^{K}\right\rangle \\
\langle j,l\left\vert \psi _{\mathrm{R}}^{K}\right\rangle%
\end{array}%
\right) =\Omega \eta _{l}e^{iKj}\left(
\begin{array}{c}
\lbrack 1-\left( -1\right) ^{l}]\left( -\kappa \right) ^{\left( l-1\right)
/2} \\
\lbrack 1+\left( -1\right) ^{l}]\left( -\kappa \right) ^{\left( M-l\right)
/2}%
\end{array}%
\right)
\end{equation}%
where $\left\vert j,l\right\rangle =c_{j,l}^{\dagger }\left\vert
0\right\rangle $\ denotes the position state and $\Omega =\sqrt{\left(
1-\kappa ^{2}\right) /4N}$. Here $\left\vert \psi _{\mathrm{L,R}%
}^{K}\right\rangle $ represents the edge state at left or right.\ The
features of the edge states are obvious: (i) Nonzero probability is only
located at the same one triangular sublattice. Then they have no chirality,
without current for any flux, which is different from the square lattice
\cite{Stanescu,Goldman,Dario}. (ii) In the case of $\kappa =0$, i.e., $\cos
\left( K_{\mathrm{c}}/2+\phi _{\mathrm{c}}\right) =0$, $\left\vert \psi _{%
\mathrm{L,R}}^{K}\right\rangle $ is reduced to the perfect edge state%
\begin{equation}
\left(
\begin{array}{c}
|\psi _{\mathrm{L}}^{K_{\mathrm{c}}}\rangle \\
|\psi _{\mathrm{R}}^{K_{\mathrm{c}}}\rangle%
\end{array}%
\right) =\sum_{j=1}^{N}\frac{e^{iK_{\mathrm{c}}j}}{\sqrt{N}}\left(
\begin{array}{c}
\left\vert j,1\right\rangle \\
\left\vert j,M\right\rangle%
\end{array}%
\right) ,
\end{equation}%
even for finite $M$, in which the particle is completely located at the
boundary.
\begin{figure}[tbp]
\centering
\par
\includegraphics[ bb=4 469 544 794, width=0.6\textwidth, clip]{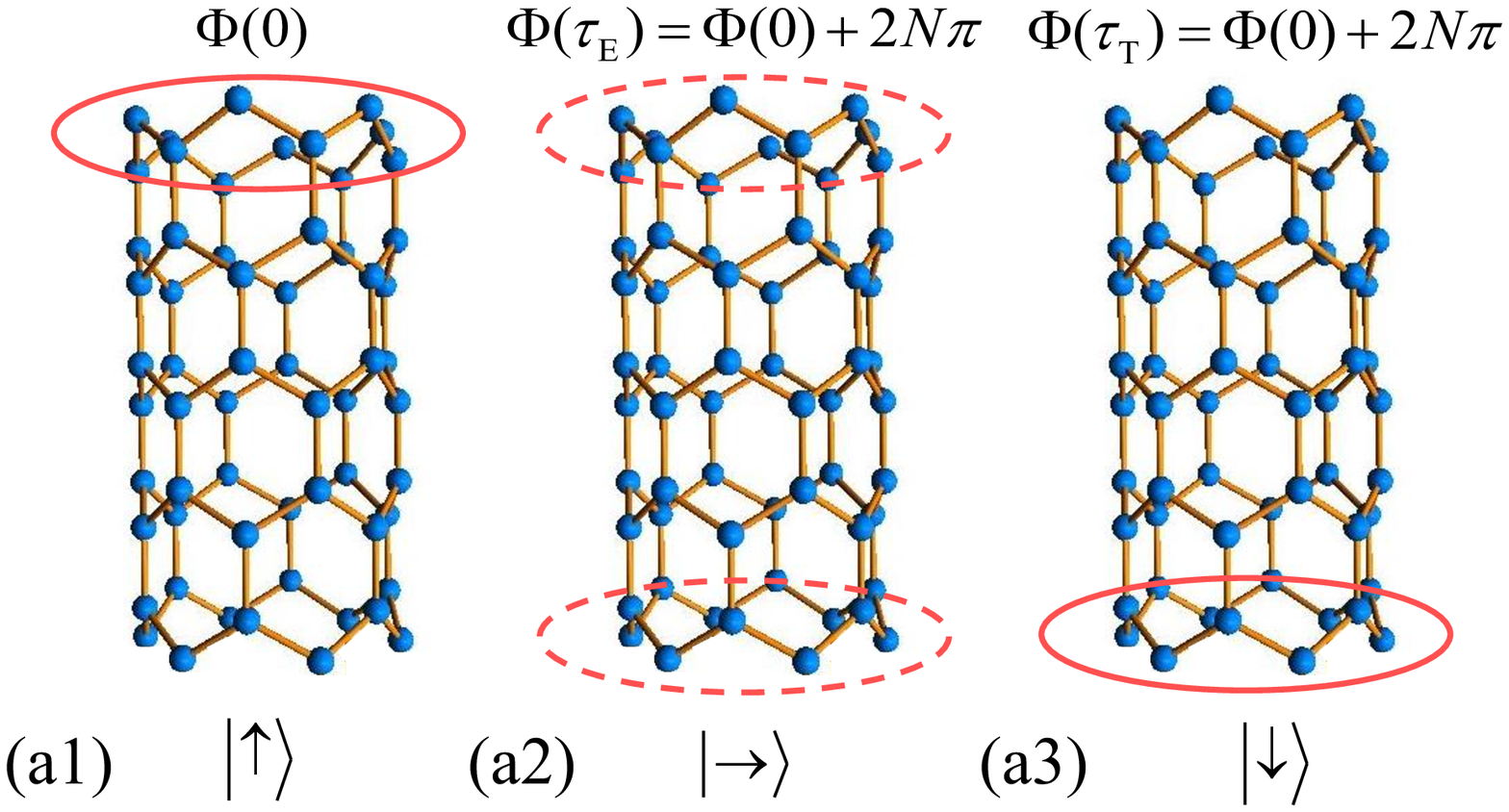} %
\subfigure{\includegraphics[bb=0 10 595 560,width=0.4\textwidth]{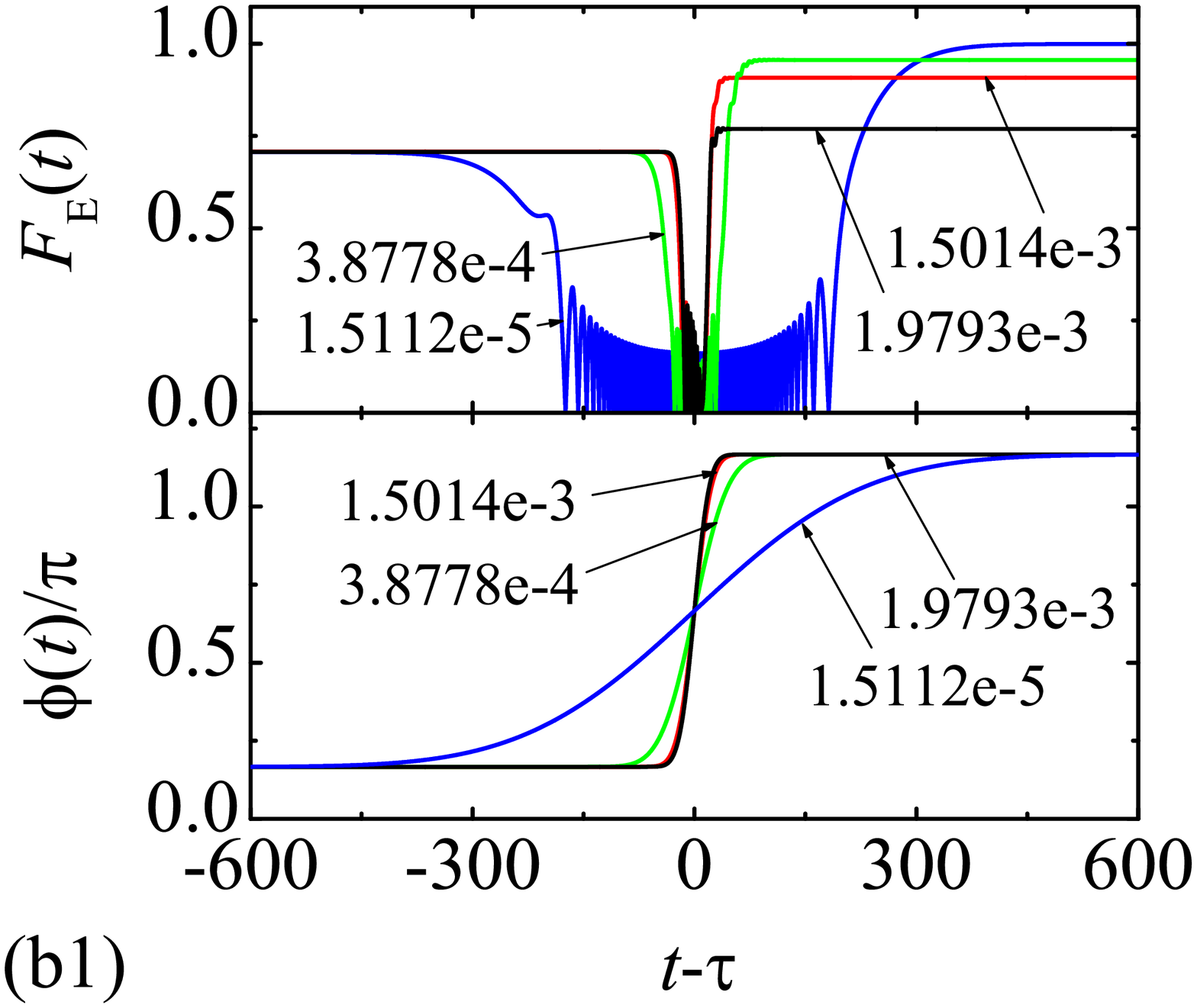}}
\subfigure{\includegraphics[bb=0 10 595 560,width=0.4\textwidth]{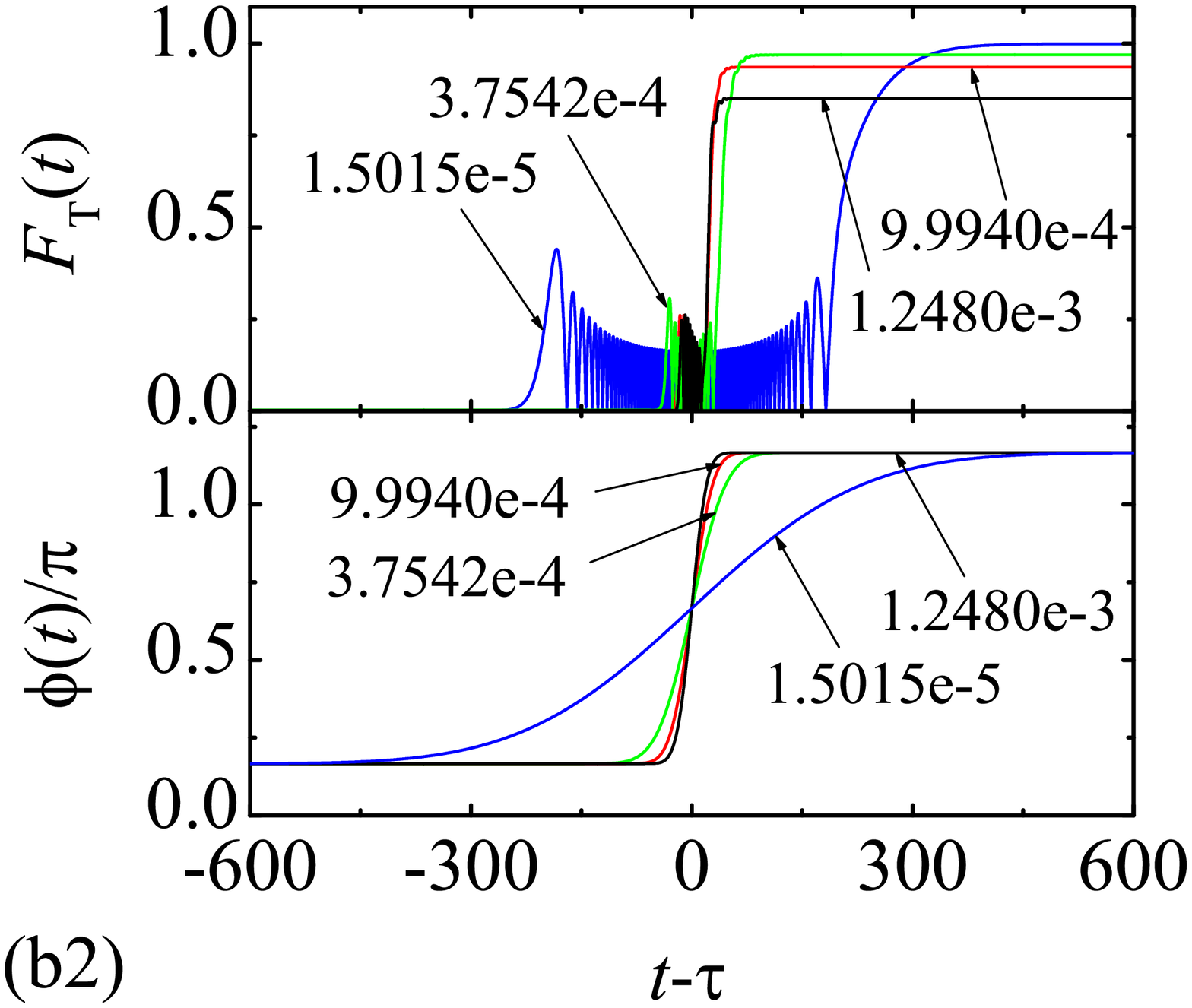}}
\par
\caption{(color online). Schematic illustration of the manipulation for a
perfect edge state in a graphene tube, which is represented by a red circle.
When varying the flux slowly, state (a1) can evolve to (a2) and (a3), where
(a2) represents a maximal entangled state. (b1) and (b2) are plots of the
fidelities under the control of Gaussian pulse fluxes with several typical $%
\protect\sigma $. Here the time $t$ is in units of $1/\Delta $. }
\label{fig4}
\end{figure}
\begin{figure}[tbp]
\centering
\includegraphics[ bb=71 238 508 535, width=0.45\textwidth, clip]{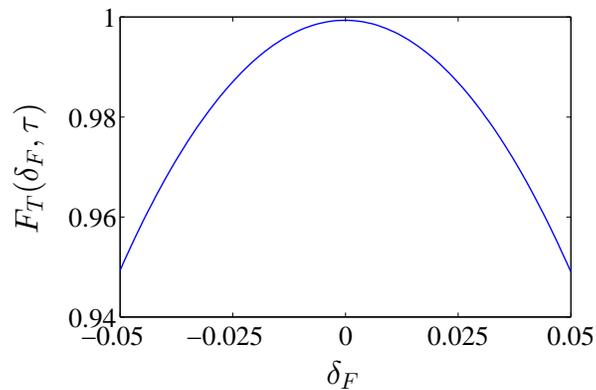}
\par
\caption{(color online). Plots of the fidelities under the control of
Gaussian pulses with several typical deviations from the standard. }
\label{fig5}
\end{figure}
Perfect edge states $|\psi _{\mathrm{L}}^{K_{\mathrm{c}}}\rangle $\ and $%
|\psi _{\mathrm{R}}^{K_{\mathrm{c}}}\rangle $\ can be regarded as the qubit
states $\left\vert \uparrow \right\rangle $ and $\left\vert \downarrow
\right\rangle $ of a nanoscale qubit protected by the gap or topology. The
qubit states can be controlled by varying the flux adiabatically. We take $%
|\psi _{\mathrm{L}}^{K_{\mathrm{c}}}\rangle $\ as an initial state, for
example. As flux changes adiabatically from $\phi _{\mathrm{c}}$, state $%
|\psi _{\mathrm{L}}^{K_{\mathrm{c}}}\rangle $\ is separated as two
instantaneous eigenstates at the edges of two bands. When\textbf{\ }$\phi
\left( \tau \right) =\phi _{\mathrm{c}}+n\pi $\textbf{\ }with $n=0,1,2,...$,
the evolved state becomes an edge state again. However, it may be the
superposition of two edge states, i.e., $c_{\mathrm{L}}|\psi _{\mathrm{L}%
}^{K_{\mathrm{c}}}\rangle +c_{\mathrm{R}}|\psi _{\mathrm{R}}^{K_{\mathrm{c}%
}}\rangle $, where the coefficients $c_{\mathrm{L}}$\ and $c_{\mathrm{R}}$\
arise from the dynamical phase, $c_{\mathrm{L}}=\cos \alpha $, $c_{\mathrm{R}%
}=i\sin \alpha $, and $\alpha =-\int\nolimits_{0}^{\tau }E_{+}\left(
t\right) \mathrm{d}t$, depending on the passage of the adiabatical process.
Here\textbf{\ }$E_{+}\left( t\right) $\ is the eigenenergy at the edge of
positive band.\textbf{\ }Proper passage allows to obtain the maximal
entangled edge state $\left\vert \rightarrow \right\rangle =(|\psi _{\mathrm{%
L}}^{K_{\mathrm{c}}}\rangle \pm i|\psi _{\mathrm{R}}^{K_{\mathrm{c}}}\rangle
)/\sqrt{2}$, or distant edge state $\left\vert \downarrow \right\rangle
=|\psi _{\mathrm{R}}^{K_{\mathrm{c}}}\rangle $.\

Here we demonstrate this process by a simplest case. We consider a graphene
tube with $N=3$ and $M=4$, which can be mapped to three $4$-site SSH chains.
Initially, we have $K_{\mathrm{c}}=2\pi /3$ and $\phi _{\mathrm{c}}=\pi /6$%
.\ Two perfect edge states are%
\begin{eqnarray}
|\mathrm{L}\rangle &=&(e^{i2\pi /3}\left\vert 1,1\right\rangle +e^{i4\pi
/3}\left\vert 2,1\right\rangle +\left\vert 3,1\right\rangle )/\sqrt{3},
\notag \\
|\mathrm{R}\rangle &=&(e^{i2\pi /3}\left\vert 1,4\right\rangle +e^{i4\pi
/3}\left\vert 2,4\right\rangle +\left\vert 3,4\right\rangle )/\sqrt{3}.
\end{eqnarray}%
When we vary the flux adiabatically from $\phi \left( 0\right) =\pi /6$\ to $%
\phi \left( \tau \right) =7\pi /6$,\ state $|\mathrm{L}\rangle $\ can evolve
to $|\mathrm{R}\rangle $ if $\alpha =-\int\nolimits_{0}^{\tau }E_{+}\left(
t\right) \mathrm{d}t=-\left( 2n-1\right) \pi /2,$\ to $\left( |\mathrm{L}%
\rangle +i|\mathrm{R}\rangle \right) /\sqrt{2}$, if $\alpha =-\left(
4n-3\right) \pi /4$, and to $\left( |\mathrm{L}\rangle -i|\mathrm{R}\rangle
\right) /\sqrt{2}$, if $\alpha =-\left( 4n-1\right) \pi /4$, where $n$ is a
large positive integer and $E_{+}\left( t\right) =\sqrt{\kappa \left(
t\right) ^{2}+1/2}-1/2$. For large $N$, the Zener tunneling occurs, reducing
the fidelity of quantum control schemes. In order to demonstrate the scheme,
we perform the numerical simulations to compute the evolved wave function%
\begin{equation}
\left\vert \Psi \left( t\right) \right\rangle =\mathcal{T}\exp
\{-i\int_{0}^{t}H[\phi (t)]\mathrm{d}t\}|\mathrm{L}\rangle
\end{equation}%
where $\phi (t)$\ takes a Gaussian form,%
\begin{equation}
\phi (t)=\phi _{\mathrm{c}}+\sqrt{\sigma \pi }\int_{0}^{t}e^{-\sigma (t-\tau
/2)^{2}}\text{\textrm{d}}t.  \label{phi_t}
\end{equation}%
We employ the fidelities
\begin{eqnarray}
F_{\mathrm{E}}\left( t\right) &=&|\left\langle \Psi \left( t\right)
\right\vert \left( |\mathrm{L}\rangle \pm i|\mathrm{R}\rangle \right) /\sqrt{%
2}|, \\
F_{\mathrm{T}}\left( t\right) &=&\left\vert \left\langle \Psi \left(
t\right) \right\vert \mathrm{R}\rangle \right\vert ,
\end{eqnarray}%
to demonstrate the two schemes. Under the assumption of adiabatical process,
$F_{\mathrm{E}}\left( t\right) $\ and $F_{\mathrm{T}}\left( t\right) $\
reach unit when $\sigma $ takes some discrete values, meeting the
restrictions for $\alpha \left( \tau \right) $\ and $\phi (\tau )$. However,
Zener tunneling will influence the fidelity as $\sigma $\ increases. We plot
the fidelities in Fig. \ref{fig4} as functions of time with several typical
values of $\sigma $ for $N=3,M=24$. We see that this scheme is achievable
with high fidelity even for the quasi-adiabatic process.

To conclude our analysis, we briefly comment on the experimental prospects
of detecting topological phase transition, edge states, and quantum state
transfer.\textbf{\ }Artificial honeycomb lattices have been designed and
fabricated in semiconductors \cite{Gibertini,DeSimoni,Singha},
molecule-by-molecule assembly \cite{Gomes}, optical lattices \cite%
{Wunsch,Soltan-Panahi,Tarruell}, and photonic crystals \cite%
{Haldane,Sepkhanov1,Sepkhanov2,Peleg}. These offer a tunable platform for
studying their topological and correlated phases. According to our analysis
above, the topological phase transition in honeycomb lattice is
equivalent to that in the SSH model. In fact, the Zak phase which is a phase
degree of freedom for the SSH model is measured in reciprocal space by using
spin-echo interferometry with ultracold atoms \cite{Atala}. As for the
detection of the perfect edge states found in our work, we propose the
following scheme. We note that a perfect zero-mode state is independent of $%
M$, i.e., it can appear in a small sized system.

We consider a graphene tube with $N=4$ and $M=4$, as an example to
illustrate the main point. In the absence of flux, there are\ two perfect
edge states
\begin{eqnarray}
|\mathrm{L}\rangle  &=&(i\left\vert 1,1\right\rangle -\left\vert
2,1\right\rangle -i\left\vert 3,1\right\rangle +\left\vert 4,1\right\rangle
)/2,  \notag \\
|\mathrm{R}\rangle  &=&(i\left\vert 1,4\right\rangle -\left\vert
2,4\right\rangle -i\left\vert 3,4\right\rangle +\left\vert 4,4\right\rangle
)/2,
\end{eqnarray}%
which have zero energy and are well separated from other levels. When a
small flux $\phi $ is switched on, the two degenerate levels are splitted as $%
\pm 4\phi ^{2}$. In the half-filled case, a small graphene tube acts as a
two-level system. This artificial atom can be detected by the absorption and
the emission of photons with frequency resonating to $8\phi ^{2}$.

For the experimental realization of quantum state transfer, the varying flux
affects the fidelity in the following two aspects in practice. (i) The
deviation of a magnetic flux from the optimal form in the adiabatic limit.
Here we consider a pulse flux as the form%
\begin{equation}
\phi (\delta _{\mathrm{F}},t)=\phi (t)+\delta _{\mathrm{F}}\sqrt{\sigma \pi }%
\int_{0}^{t}e^{-\sigma (t-\tau /2)^{2}}\text{\textrm{d}}t,
\end{equation}%
where $\delta _{\mathrm{F}}$\ is introduced as a quantity to express the
strength of deviation. Fig. \ref{fig5} is the plots of the fidelities under
the control of Gaussian pulse fluxes $\phi (\delta _{\mathrm{F}},t)$, which are obtained by numerical simulations for several typical values of $\delta _{%
\mathrm{F}}$. It indicates that the deviation of the flux would reduce the
fidelity for the process of adiabatic transfer. And we believe that a similar
conclusion could also be achieved from the process of generating the maximal
entanglement. (ii) The speed of varying flux. The adiabatic process requires a sufficiently slow speed of the flux. A fast pulse would induce to Zener tunneling, reducing the fidelity. From Fig. \ref{fig4}, we can see the influence of the
speed on the fidelity, which provides a theoretical estimation for
experiments.

\section*{Discussion}

We have proposed two ways to rewrite the Hamiltonian of a flux-threaded
graphene tube by the sum of several independent sub-Hamiltonians. Each of
them represents a system that may have nontrivial topology or not, which can
be determined by the threading flux.\textbf{\ }For an open tube, such
topology emerges as zero-mode edge states according to the bulk-boundary
correspondence, which is still controllable by the flux. In addition, we
have shown the existence of the perfect edge state even for a small length
tube. Such kind of states can be transferred and entangled by adiabatically
varying the flux, which has the potential application to design a nanoscale
quantum device. There are three advantages as a quantum device: (i) The
perfect edge states of two ends are well distinguishable, pointer states,
even for a small length tube, (ii) Midgap states are well protected by the
gap, (iii) Owing to the finite size effect, the gap between the controlled
levels and others is finite for any flux, suppressing the Zener tunneling
and allowing the realization of an adiabatic process.

\section*{Methods}

\textbf{Second equivalent Hamiltonian.} In this section, we investigate the
topological quantum phase transition in a graphene tube and propose an
implementation on the transfer and entanglement of perfect edge states
driven by a threading magnetic field. The notes here provide additional
details on the derivations of an alternative geometrical representation of
the system with periodic boundary condition and the wave functions of edge
states for the system with open boundary condition. A demonstration of
perfect edge states and the processes of state transfer and entanglement
generation is presented.

\begin{figure}[tbp]
\centering
\includegraphics[ bb=0 352 538 605, width=0.75\textwidth, clip]{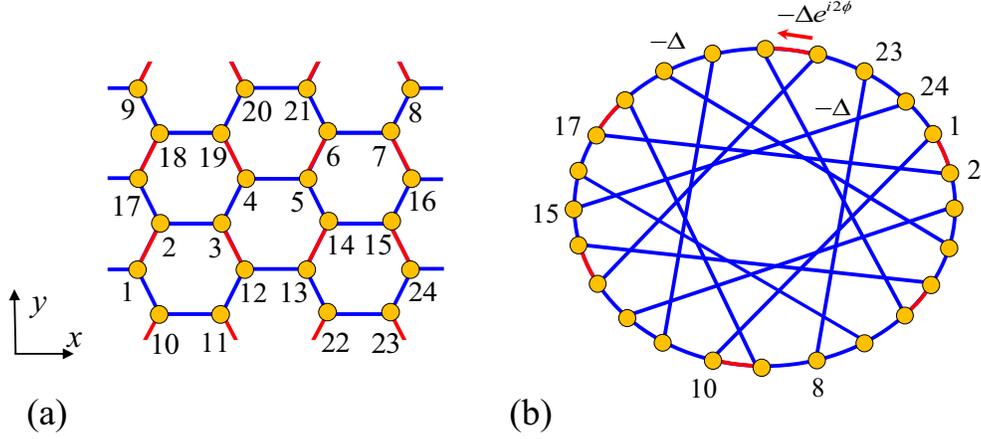}
\par
\caption{(color online). (a) Sketch of a graphene tube with periodic
boundary condition for $N=3$, $M=8$. The lattice is indexed in another way.
(b) Based on this method, the original lattice can be regarded as an $24$%
-site ring with long-range hoppings.}
\label{fig6}
\end{figure}

In order to demonstrate the obtained winding numbers in Eq. (\ref{WN}) are
topological invariant, we consider another way to exhibit the topological
feature of the honeycomb tube. We index the lattice in an alternative way,
which is illustrated in Fig. \ref{fig6} for the case with $N=3$, $M=8$. Now
we consider an $N_{0}$-site\ tube lattice with $N_{0}=MN$, where $M$ and $N$
satisfy the condition $C\times N+1=D\times B$. Here $B=M/4-N\left[ M/\left(
4N\right) \right] $,\ $C$\ and $D$ are the minimum non-negative integers
that meet the condition. For example, taking $N=3$, $M$\ can be taken as $%
12m-4$ ($B=2$, $C=1$, $D=2,m\geq 1$) while taking $N=4$, $M$\ can be taken
as $16m-12$ ($B=1$, $C=0$, $D=1,m\geq 1$). For finite $N$, $M$\ can be
infinite, leading to a thermodynamic limit system, which shares the same
property as that mentioned in the former sections. Such an arrangement
allows us to rewrite the Hamiltonian in the form
\begin{equation}
H=-\sum_{n=1}^{N_{0}}c_{n}^{\dagger }c_{n+1}-\sum_{n=1}^{N_{0}/4}\left(
e^{i2\phi }-1\right) c_{4n-3}^{\dagger }c_{4n-2}-\sum_{n=1}^{N_{0}/4}\left(
e^{i2\phi }c_{4n}^{\dagger }c_{4n+DM-1}+c_{4n-2}^{\dagger
}c_{4n+DM-3}\right) +\mathrm{H.c.},  \label{H_eq}
\end{equation}%
with the periodic boundary condition $c_{n+N_{0}}^{\dag }=c_{n}^{\dag }$. It
represents an $N_{0}$-site ring with long-range hoppings. Fig. \ref{fig6}
schematically illustrates the lattice structure. Being the same system, we
want to examine what topology is hidden in the new Hamiltonian, although we
believe that it should be invariant. To this end, we introduce alternative
Fourier transformations%
\begin{equation}
\left\{
\begin{array}{c}
a_{1,k}=\frac{e^{-i\phi /2}}{\sqrt{N_{0}/4}}\sum_{j}^{N_{0}/4}e^{-ik\left(
j-1/4\right) }c_{4j-3}, \\
a_{2,k}=\frac{e^{i\phi /2}}{\sqrt{N_{0}/4}}\sum_{j}^{N_{0}/4}e^{-ik\left(
j-1/4\right) }c_{4j-2}, \\
a_{3,k}=\frac{e^{i\phi /2}}{\sqrt{N_{0}/4}}\sum_{j}^{N_{0}/4}e^{-ik\left(
j+1/4\right) }c_{4j-1}, \\
a_{4,k}=\frac{e^{-i\phi /2}}{\sqrt{N_{0}/4}}\sum_{j}^{N_{0}/4}e^{-ik\left(
j+1/4\right) }c_{4j},%
\end{array}%
\right.
\end{equation}%
where $k=8\pi l/N_{0}$, ($l\in \lbrack 1,N_{0}/4]$). Since the translational
symmetry of the Hamiltonian (\ref{H_eq}), we still have $H=\sum_{k}H^{k}$
with $[H^{k},H^{k^{\prime }}]=0$, where%
\begin{equation}
H^{k}=-[e^{i\phi }+e^{-i\left( DMk/4+\phi \right)
}]\sum_{j=1}^{2}a_{2j-1,k}^{\dag
}a_{2j,k}-e^{ik/2}\sum_{j=1}^{2}a_{2j,k}^{\dag }a_{2j+1,k}+\mathrm{H.c.},
\end{equation}%
with the periodic boundary $a_{5,k}=a_{1,k}$. Similarly, we have the
corresponding pseudo-spin representation%
\begin{equation}
H^{k}=2\sum_{\lambda =\pm }\vec{r}_{\lambda }\left( k\right) \cdot \vec{s}%
_{\lambda ,k},
\end{equation}%
where $\vec{s}_{\lambda ,k}$ ($\lambda =\pm $) is defined as pseudo spin
operators%
\begin{eqnarray}
s_{\lambda ,k}^{+} &=&(s_{\lambda ,k}^{-})^{\dag }=a_{\lambda ,k}^{\dag
}b_{\lambda ,k}, \\
s_{\lambda ,k}^{z} &=&\frac{1}{2}(a_{\lambda ,k}^{\dag }a_{\lambda
,k}-b_{\lambda ,k}^{\dag }b_{\lambda ,k}),
\end{eqnarray}%
which obey%
\begin{equation}
\lbrack s_{\lambda ,k}^{+},s_{\lambda ^{^{\prime }},k}^{-}]=2s_{\lambda
,k}^{z}\delta _{\lambda ,\lambda ^{^{\prime }}},[s_{\lambda
,k}^{z},s_{\lambda ^{^{\prime }},k}^{\pm }]=\pm s_{\lambda ,k}^{\pm }\delta
_{\lambda ,\lambda ^{^{\prime }}},
\end{equation}%
and here $a_{\lambda ,k}$ and $b_{\lambda ,k}$ are defined as%
\begin{equation}
\left\{
\begin{array}{c}
a_{\lambda ,k}=\frac{1}{\sqrt{2}}\sum_{j=1}^{2}\left( -\lambda \right)
^{j}a_{2j-1,k}, \\
b_{\lambda ,k}=\frac{1}{\sqrt{2}}\sum_{j=1}^{2}\left( -\lambda \right)
^{j}a_{2j,k}.%
\end{array}%
\right.
\end{equation}%
The field $\vec{r}_{\lambda }\left( k\right) =\left( x_{\lambda }\left(
k\right) ,y_{\lambda }\left( k\right) \right) $\ is%
\begin{equation}
\left\{
\begin{array}{c}
x_{\lambda }\left( k\right) =-\cos \phi +\lambda \cos \left( k/2\right)
-\cos \left( DMk/4+\phi \right) \\
y_{\lambda }\left( k\right) =\sin \phi +\lambda \sin \left( k/2\right) -\sin
\left( DMk/4+\phi \right)%
\end{array}%
\right. .
\end{equation}

\begin{figure}[tbp]
\centering
\includegraphics[ bb=1 164 541 765, width=0.45\textwidth, clip]{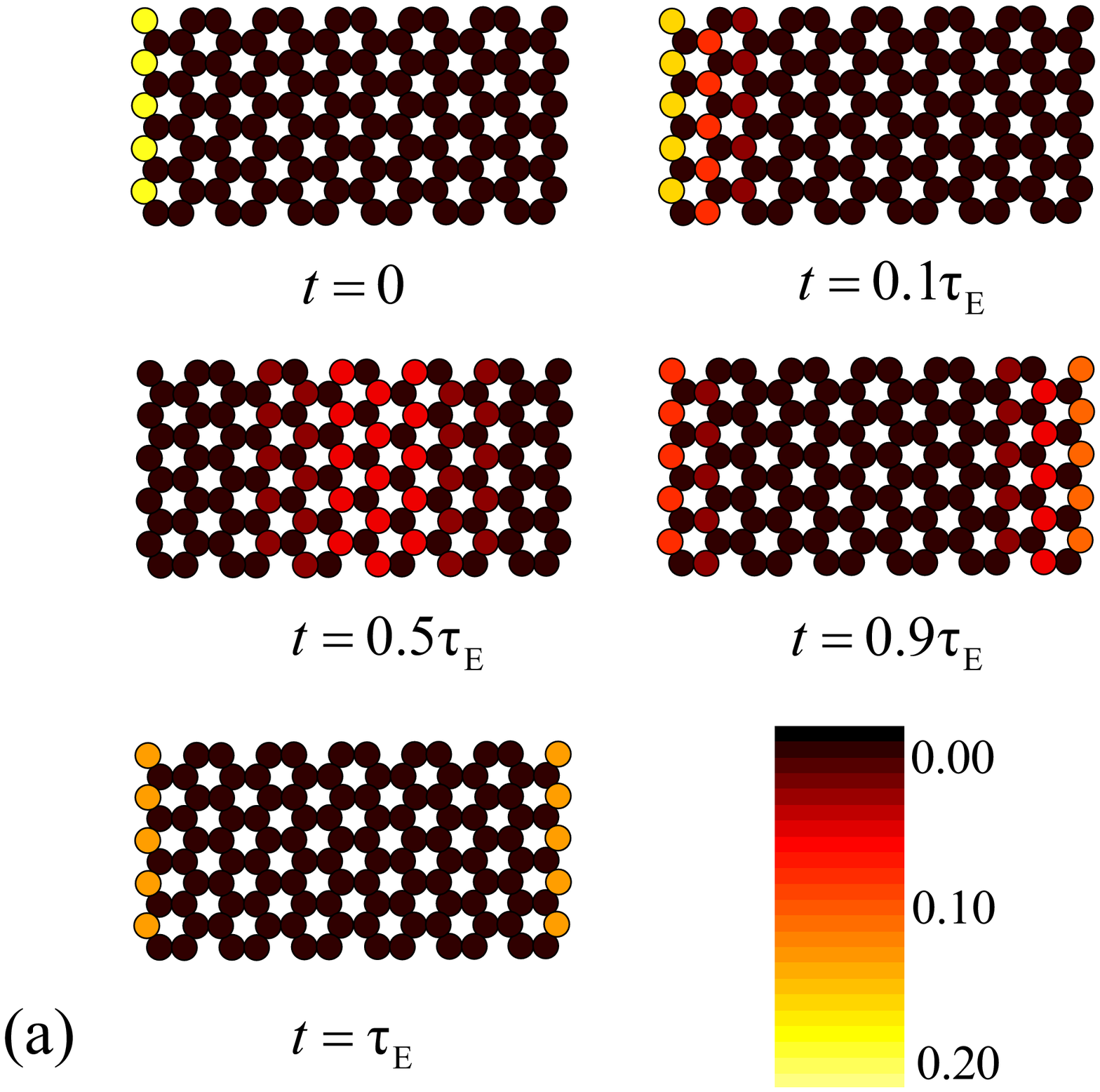} %
\includegraphics[ bb=1 164 541 785, width=0.45\textwidth, clip]{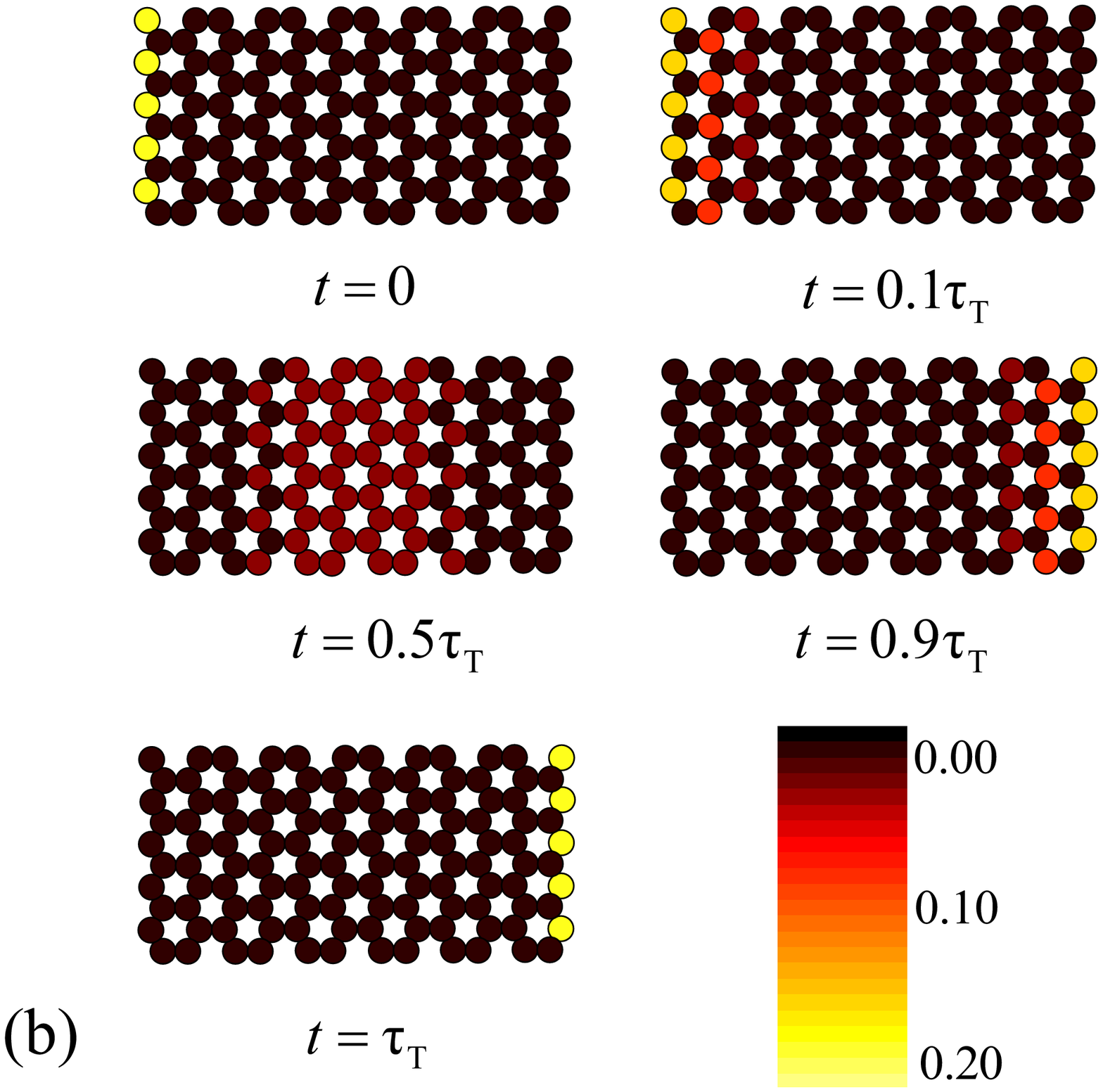}
\par
\caption{(color online). The stroboscopic picture of the probability
distribution function for a perfect edge state and the dynamic process as
flux varies in the graphene tube with $N=5,M=24,$ and $K_{c}=4\protect\pi /5$%
, obtained by numerical simulations.\ The initial state is a perfect edge
state with all probability completely distributes at the $5$ leftmost sites.
(a) Generation of a maximal entangled edge state with duration time $\protect%
\tau _{\mathrm{E}}=$ $8.9760\times 10^{3}\Delta ^{-1}$. (b) Transfer of edge
state with $\protect\tau _{\mathrm{T}}$ $=6.2831\times 10^{3}\Delta ^{-1}$. }
\label{fig7}
\end{figure}
We are interested in the geometry of the curve represented by the above
parameter equations. We note that%
\begin{eqnarray}
x_{\lambda }\left( k+4\pi \right) &=&x_{\lambda }\left( k\right) , \\
y_{\lambda }\left( k+4\pi \right) &=&y_{\lambda }\left( k\right) ,  \notag
\end{eqnarray}%
and%
\begin{eqnarray}
x_{-}\left( k\right) &=&x_{+}\left( k+2\pi \right) , \\
y_{-}\left( k\right) &=&y_{+}\left( k+2\pi \right) ,  \notag
\end{eqnarray}%
which allow us to rewrite the Hamiltonian as%
\begin{equation}
H=2\sum_{k}\vec{r}\left( k\right) \cdot \vec{s}_{k},
\end{equation}%
where the field and pseudo-spin operator are redefined as%
\begin{eqnarray}
\vec{r}\left( k\right) &=&\vec{r}_{+}\left( k\right) ,k\in \left( 0,4\pi %
\right] , \\
\vec{s}_{k} &=&\left\{
\begin{array}{cc}
\vec{s}_{+,k}, & k\in \left( 0,2\pi \right] \\
\vec{s}_{-,k-2\pi }, & k\in (2\pi ,4\pi ]%
\end{array}%
\right. .
\end{eqnarray}%
The equivalent Hamiltonian (\ref{H_eq}) provides another platform to study
the geometry of the band. We find that $\vec{r}\left( k\right) $ is not
smooth as $k$ increases continuously from $0$ to $4\pi $, which indicates
that the curve $\left\{ \vec{r}\left( k\right) \right\} $ consists of
several independent loops. Actually, considering discrete $k$, we can
decompose the set of $k$ into $N$ groups by dividing $l$ into $l_{n}=mN+n$
with $m\in \left[ 0,M-1\right] $, $n\in \left[ 1,N\right] $. Then we have
\begin{equation}
k_{n}=\left\{
\begin{array}{cc}
8\pi l_{n}/N_{0}, & k_{n}\in \left( 0,2\pi \right] \\
2\pi +8\pi l_{n}/N_{0}, & k_{n}\in (2\pi ,4\pi ]%
\end{array}%
\right. .
\end{equation}%
Noting $DMk_{n}/4=2\pi D\left( m+n/N\right) $, the parameter equation of the
$n$th\ curve becomes%
\begin{equation}
\left\{
\begin{array}{c}
x\left( k_{n}\right) =-\cos \left( \phi \right) +\cos \left( k_{n}/2\right)
-\cos \left( 2\pi nD/N+\phi \right) \\
y\left( k_{n}\right) =\sin \left( \phi \right) +\sin \left( k_{n}/2\right)
-\sin \left( 2\pi nD/N+\phi \right)%
\end{array}%
\right. ,
\end{equation}%
which could be rewritten as
\begin{equation}
\left\{
\begin{array}{c}
x\left( k_{n}\right) =x_{0}^{n}+\cos \left( k_{n}/2\right) \\
y\left( k_{n}\right) =y_{0}^{n}+\sin \left( k_{n}/2\right)%
\end{array}%
\right. ,
\end{equation}%
where
\begin{equation}
\left\{
\begin{array}{c}
x_{0}^{n}=x_{\mathrm{c}}-\cos \left( 2\pi nD/N+\phi \right) , \\
y_{0}^{n}=y_{\mathrm{c}}-\sin \left( 2\pi nD/N+\phi \right) .%
\end{array}%
\right.
\end{equation}%
with%
\begin{equation}
\left\{
\begin{array}{c}
x_{\mathrm{c}}=-\cos \left( \phi \right) , \\
y_{\mathrm{c}}=\sin \left( \phi \right) .%
\end{array}%
\right.
\end{equation}%
The geometry of the curves are obvious, representing $N$\ circles of unit
radius with the center $\left( x_{0}^{n},y_{0}^{n}\right) $. Furthermore,
the positions of circle center are simply characterized as $\left(
x_{0}^{n},y_{0}^{n}\right) $. These circle centers would contribute a $N$
regular polygon and be concyclic points of a circumcircle with the
circumcenter $\left( x_{\mathrm{c}},y_{\mathrm{c}}\right) $\ and the unit
circumradius.

\vspace{2ex}\textbf{Bethe ansatz of Edge states.} Now we turn to the
derivation of the wave functions of the edge states for infinite-$M$ system.
In the single-particle invariant subspace, the Hamiltonian of Eq. (\ref{H_K2}%
) with open boundary is written as
\begin{equation}
\underline{H}_{K}=-\kappa \sum_{m=1}^{M/2}\underline{\left\vert
2m-1\right\rangle }\underline{\left\langle 2m\right\vert }-\sum_{m=1}^{M/2}%
\underline{\left\vert 2m\right\rangle }\underline{\left\langle
2m+1\right\vert }+\mathrm{H.c.},  \label{HK_Ansatz}
\end{equation}%
where $\underline{\left\vert l\right\rangle }=a_{K,l}^{\dag }\left\vert
0\right\rangle $ is a position state on the SSH chain. We are interested in
the edge state, which corresponds to the bound state at the ends for
infinite $M$. We focus on the bound state at the end of small $m$, then the
other one with largest $m$ is straightforward.

The Bethe ansatz wave function has the form%
\begin{equation}
\underline{\left\vert \psi \right\rangle }=\sum_{m=1}^{\infty }(e^{-\beta
\left( 2m-1\right) }\underline{\left\vert 2m-1\right\rangle }+De^{-\beta
\left( 2m\right) }\underline{\left\vert 2m\right\rangle }),
\end{equation}%
and the Schrodinger equation $\underline{H}_{K}\underline{\left\vert \psi
\right\rangle }=E\underline{\left\vert \psi \right\rangle }$ gives%
\begin{equation}
\left\{
\begin{array}{c}
-\kappa De^{-\beta }=E, \\
-\kappa e^{\beta }-e^{-\beta }=ED, \\
-De^{2\beta }-\kappa D=Ee^{\beta }.%
\end{array}%
\right.
\end{equation}%
Solving these equations, we have%
\begin{equation}
D=0,e^{2\beta }=-1/\kappa ,E=0,
\end{equation}%
in the case of $0<\left\vert \kappa \right\vert <1$. Then\ the normalized
bound state wave function yields%
\begin{equation}
\underline{\left\vert \psi \right\rangle }=\sqrt{1-\kappa ^{2}}%
\sum_{m=1}^{\infty }\left( -\kappa \right) ^{m-1}\underline{\left\vert
2m-1\right\rangle }.  \label{BS WF}
\end{equation}%
While, in the case of $\kappa =0$, from the Hamiltonian (\ref{HK_Ansatz}) it
is easy to check that there exist a perfect edge state $\underline{%
\left\vert 1\right\rangle }$\ and two types of highly degenerate eigenstates
$\left( \underline{\left\vert 2m\right\rangle }\pm \underline{\left\vert
2m+1\right\rangle }\right) /\sqrt{2}$\ with their eigenenergies $\mp 1$\
(here $m\geq 1$). To demonstrate the processes of perfect transfer and
entanglement generation for a perfect edge state, we plot the profile of
probability distributions at several typical time in Fig. \ref{fig7}. The
results are obtained by exact diagonalization.


\section*{Acknowledgement}

We acknowledge the support of the National Basic Research Program (973
Program) of China under Grant No. 2012CB921900 and CNSF (Grant No. 11374163).

\section*{Author contributions}

S.L. did the derivations and edited the manuscript, Z.S. conceived the
project and drafted the manuscript. G.Z. and C.L. checked the results and
revised the manuscript. All authors discussed the results and reviewed the
manuscript.

\section*{Additional information}

The authors do not have competing financial interests.

\end{document}